# Information vs. Uncertainty as the Foundation for a Science of Environmental Modeling

Grey S. Nearing[1,2,3] & Hoshin V. Gupta[4]

[1]*NASA Goddard Space Flight Center, Hydrological Sciences Laboratory; Greenbelt, MD; grey.s.nearing@nasa.gov*
[2]*National Center for Atmospheric Research, Research Applications Laboratory; Boulder, CO; grey@ucar.edu*
[3]*University of Maryland Baltimore County, Department of Computer Science and Electrical Engineering; Catonsville, MD; grey@umbc.edu*
[4]*University of Arizona, Department of Hydrology and Atmospheric Sciences; Tucson, AZ; hoshin@email.arizona.edu*

**Abstract:** Information accounting provides a better foundation for hypothesis testing than does uncertainty quantification. A quantitative account of science is derived under this perspective that alleviates the need for epistemic bridge principles, solves the problem of ad hoc falsification criteria, and deals with verisimilitude by facilitating a general approach to process-level diagnostics. Our argument is that the well-known inconsistencies of both Bayesian and classical statistical hypothesis tests are due to the fact that probability theory is an insufficient logic of science. Information theory, as an extension of probability theory, is required to provide a complete logic on which to base quantitative theories of empirical learning. The organizing question in this case becomes not whether our theories or models are more or less true, or about how much uncertainty is associated with a particular model, but instead whether there is any information available from experimental data that might allow us to improve the model. This becomes a formal hypothesis test, provides a theory of model diagnostics, and suggests a new approach to building dynamical systems models.

1. **Motivation and Overview**

The ostensible objective of this paper is to outline a conceptual framework for testing and evaluating models in a way that facilitates reliable scientific learning. An important example of this is in the Earth and Environmental Sciences, where improved process understanding is critical to making informed decisions under nonstationary climatological and anthropogenic conditions. Put simply, complex systems models must be able to produce accurate and reliable predictions due to the fact that they are in some sense isomorphic with the systems that they represent, rather than because they are developed or calibrated to agree with observations.

An optimistic perspective might be that our understanding of how to evaluate complex systems models is in a pre-normal phase. There are effectively innumerable reports, published across all fields of science, that propose and/or apply varied and non-commensurate methods for model evaluation, benchmarking, intercomparison, etc. Oreskes & Belitz (2001) recognize a fundamentally subjective component to model evaluation. Related to the Environmental Sciences in particular, Beck et al. (2009) outlined in their NSF white paper a 'Grand Challenge' related to the need for *"radically novel procedures and algorithms … to rectify the chronic, historical deficit of the past four decades in engaging complex models systematically and successfully with field data for the purposes of learning and discovery."*

A more pessimistic view comes from recognizing that the problem of reconciling complex systems models with observations is *fundamentally* hard. The confirmation holism problem (Severo, 2012), which is an aspect of the Duhem-Quine thesis (Harding, 1976), proposes that scientists are only able to test models *in toto*, and that it is impossible to test individual scientific hypotheses. Lenhard & Winsberg (2011) argued that a consequence of this is that we *"are likely to continue to be unable to attribute the various sources of [climate model] successes and failures to their internal modeling assumptions."* Leaving aside whether the situation is quite so hopeless, it is certainly true that we currently lack any fundamental theory of how to diagnose individual components of reductionist-type models of systems that are composed of many interacting components and processes.

We are optimists. The remainder of this essay will argue that meaningful and general solutions to the related problems of *model evaluation* and *model diagnostics* require rather fundamental adjustment to both the philosophy and practice of model-based science. Our argument is that many of the practical challenges that modelers face on a day-to-day basis are due to certain deeply rooted logical inconsistencies (*e.g.,* Nearing et al., 2016b) in our standard suite of quantitative methods of inference. These inconsistencies turn out to be largely related to the fact that all of our statistical methods are built fundamentally around a concept of uncertainty (imprecise doxastic states) and that uncertainty is strictly non-quantifiable (*i.e.,* all probability distributions are wrong).

We propose that it is possible to develop a coherent method of science that does not require explicit treatment of uncertainty. Instead of using methods of inference that treat uncertainty (*e.g.,* probabilistic or statistical hypothesis tests), we base our methods of inference on measures of information. In particular, we measure whether observation data contains information that may be used to improve a model. We do this by first noting that probability theory lacks the ability to describe a certain fundamental component of epistemic behavior during empirical inquiry, and therefore cannot be a complete logic of science. This deficiency is remedied with information theory.

The basic insight, which we adopt from Jaynes (2003; p21), is that any logic of hypothesis testing acts on *statements about what we know about states of affairs* in the world (epistemic propositions), rather than on *statements about states of affairs in the world* (ontic propositions). Notice that this is exactly what occurs under all types of probabilistic inference – probabilities are epistemic statements, and these are the quantities that are manipulated during a hypothesis test or application of Bayes' theorem.

Accordingly, if predictions derived from scientific hypotheses are *statements about states of affairs* (*i.e.,* the model makes a prediction about what will happen), then any system of logic capable of supporting a scientific method must necessarily include a theory for relating such ontic propositions with epistemic propositions (*i.e.,* what we *learn* from predictive models). This effectively means assigning probabilities to model predictions and also to observation data. We will refer to methods for assigning quantitative translations between ontic and epistemic propositions as *epistemic bridges*[1]. Epistemic bridges are things like error functions, performance metrics, and/or likelihood functions. The challenge – as we will discuss in **Section 2** and associated Appendices – is that ad hoc assignment of epistemic bridges can cause substantial errors during inference (*e.g.,* Beven et al., 2008).

This problem is resolved by noting that there exists a fully developed logical theory that relates ontic with epistemic propositions. This theory is not a bridge principle – it is a necessary and unique consequence of the calculitic system of logic that supports and includes probability theory. This is described in accessible detail by (Knuth, 2004), and taking this fact seriously allows us to avoid the need for any ad hoc metrics or any choice of adequacy or falsification criteria. **Section 3** discusses how improving the underlying *logic* of hypothesis testing allows us to resolve at least many of the hard *practical* problems related to hypothesis testing and model evaluation.

We offer no opinion on scientific realism, however we do point out that any quantitative method for assigning degrees of belief about truth-values of models will necessarily contain contradiction simply because all models are approximations. And so we see it as inevitable that any formal and quantitative account of inference will not treat models as truth-apt (Bayesianism, for example, does assign doxastic measures to models, but fails to do so reliably). Because we have given up any pretense of searching for truth-values of models, it is therefore necessary to somehow draw a distinction between models that are empirically justifiable, and those that are somehow isomorphic with real-world systems. **Section 4** applies the theory from **Section 3** to this purpose.

In the first sentence of this essay we said that our *ostensible* objective is to outline a conceptual framework for evaluating models. However, there is no way to test any hypothesis except by first embedding it into one or more predictive models (Cartwright, 1983). This means that any theory of model evaluation is actually a theory of science writ large, and any theory of science is a theory of model evaluation – there is no distinction between these two things. Thus, the real objective of this essay is to propose a theory of science. Specifically, while it is true that there are no criteria (*e.g.,* falsifiability) sufficient to differentiate a hypothesis as either scientific or non-scientific (Laudan, 1983), we *can* demarcate combinations of *models plus experiment* as either scientific or non-scientific depending on whether the experiment is sufficient to indicate (either formally or informally) the information content of the hypothesis about the experimental data. Information content of a model is defined here as the fractional ability of model predictions to inform variability in experimental data.

---

[1] The term *bridge principle* refers to any explicit connection between different bodies of theory. Here we use the term *epistemic bridge* to refer to a proposition that relates any ontological theory with any epistemological theory.

## 2. Uncertainty
### 2.1. Epistemological Foundations: Logic and Probability Theory

Before we can undertake any meaningful investigation, scientific or otherwise, it is necessary to have a clearly defined logical environment. As Quine (1986; chapter 6) put it, changing our logic is akin to changing the question. Without a well-specified system of logic, we cannot ask a well-formed question and there is no hope of consistency. As alluded to above, our argument is that many of the scientific community's practical challenges related to model evaluation are actually due to primitive logical inconsistencies (Nearing et al., 2016b). Any scientific method or practice that we derive will depend completely on our logical primitives.

We will use here what is unarguably the most common system of logic, based on the axioms that Russell (1912; chapter VII) – perhaps controversially (*e.g.,* Bueno and Colyvan, 2004) – called self-evident. In this context, all propositions are either true or false (excluded middle) but not both (non-contradiction), and although we often do not know the truth-values of a given set of propositions, we may have some ability to formulate consistent doxastic states.

This system of logic, which we will hereafter refer to as the *classical logic*, leads to a standard propositional algebra[2], however this logic remains insufficient to describe any dynamic learning process because it does not contain a variable that is subject to change. To describe a dynamic learning process we need a calculus, and Cox (1946) laid the groundwork for a demonstration (Terenin and Draper, 2015) that the only scalar calculus that is consistent with the classical logic is probability theory, at least under certain relatively weak assumptions (Van Horn, 2003). For an accessible treatment of this see Jaynes (2003; chapter 1). While there certainly exist other epistemic calculi that do not derive from the classical logic (*e.g.,* Kosko, 1990, Chang, 1958, Rescher, 1968), it is necessary that we have *some* logical calculus if we want to formalize any method of science, and for the rest of this essay we will work under the classical logic and its unique scalar extension into the probability calculus.

Like all formal systems of logic, the probability calculus describes relationships between propositions. Note two things: 1) that probability distributions are always conditional on a priori information (Jaynes, 2003; p43) – at the very least a formal and explicit statement of a set of competing propositions – and 2) that all probabilities are conditional must be explicit in any coherent formulation of scientific inference (Howson, 2000; p239). In other words, probability distributions are – fundamentally – expressions of the logical doxastic consequences of available information.

It is also important to understand that although probability distributions are expressions of uncertainty, in that they describe relative doxastic states about the truth-values of various competing propositions, probability distributions cannot be related to actual uncertainty in any systematic way. This problem is well known (e.g., Knight, 1921, Taleb, 2010, Beven, 2016). The difference between having an expression of uncertainty and having an expression of uncertainty that can be related to real uncertainty is critical, and the remainder of **Section 2** outlines why our inability to estimate uncertainty means that probability theory is insufficient to support a coherent account of science. **Section 3** then describes how information theory fills the gap, and how this solves many of our day-to-day practical problems related to testing models.

### 2.2. Types of Uncertainty that Arise in Model/Hypothesis Testing

---
[2]What we call the *propositional algebra* is more often called a *propositional calculus*. It is, however, not a calculus since it does not contain any dynamic variable. It is actually an algebra, and we will refer to it as such so as to differentiate between this and a truly calculitic logic like probability theory.

To see why uncertainty is not a useful concept in science we must first understand what a model-based experiment looks like. An example of a common model-based experiment in the geosciences is illustrated in **Figure 1**. Here our model consists of various process representations denoted by $\hbar_*$. These are Hempel & Oppenheim's (1948) *explanantia*, and are statements that describe the dynamics of both the system under study and the measurement devices used to collect experimental data. The experiment also includes statements about various aspects of the system that are observable (at least in principle); these are denoted by random variables (*e.g.,* $u$, $\theta$, $y$), which we refer to as *phenomenological*[3]. Finally, the experiment includes data that contain information about some of these phenomenological components.

Note that data, represented here as $z_*$, are actually ontic in nature; for example, these are often the binary states of some computer memory. Data must be related to phenomenology via explanantia that represent a measurement processes, and it is important to remember that all measuring processes – probably including human perception and cognition (*e.g,.* Tononi, 2011) – are just dynamic physical systems. It is only after this model building process – *i.e.,* after assigning one or more phenomenological variables and measurement models – that data facilitate *statements about states of affairs*. Such statements are artifacts of models, not of data themselves. To put this another way, recognizing a fundamental separation between phenomenological model building vs. hypothesis-driven model building (*e.g.,* Chalmers, 2013; chapter 1) seems to be an error. In other words, we take Quine's (1951) all-inclusive epistemic fabric quite seriously, except – as pointed out above and by Quine (1986) – to note that we must have some system of logic to even formulate a question.

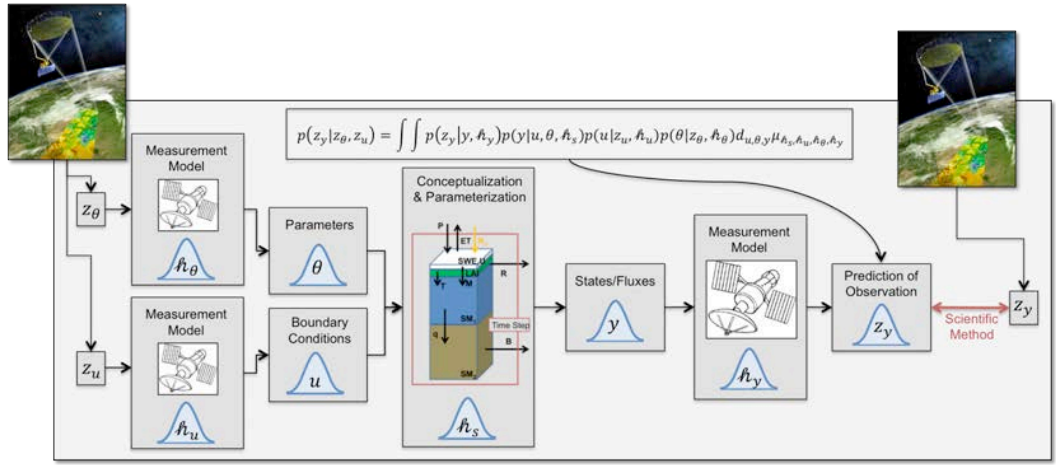

*Figure 1: A probabilistic science experiment involving a complex systems model. Process components (explanantia) are labeled $\hbar_*$, phenomenological components – those that are observable – are labeled by random variables $u, \theta, y$, and experimental data are labeled $z_*$.*

The particular example of an experiment in **Figure 1** recognizes three types of phenomenological components – model parameters $\theta$, boundary conditions $u$, and system responses $y$. More generally, any science experiment includes some experimental inputs (here $\theta$ and $u$) simply because by definition there are no accessible closed systems, and an experiment proceeds by developing and testing explanantia to predict responses of a perturbed system (perturbations may be natural or induced). We must keep track of the perturbations, in the form of information stored in data (here $z_\theta$ and $z_u$), and this information can only be interpreted through some measurement model that relates data with various phenomenological components, and it is the act of interpreting data through a measurement model that admits something like *observation error*, which is actually a form of model error. Specifically, what we typically call *measurement error* is actually error in a measurement model. It is important that we aren't tempted to reify the idea that data themselves contain error. Data are just physical manifestations of portions of the universe, and the best the scientist can do is to accurately and precisely predict their interactions with experimental data using a descriptive (and predictive) model.

---

[3]The term *phenomenological* is used to refer to things that are, in principle, observable. Whether or not an entity is in principle observable is completely an aspect of the model itself. We make no a priori (model independent) claims about observability; the theory itself that is embedded in the model tells us whether a particular entity of a particular model is at least potentially observable, either directly or indirectly.

An experiment contains two primary types of uncertainty: uncertainty related to data and uncertainty related to explanantia (various components of the model). Related to data, experimental perturbation data will generally not contain sufficient information to fully determine experimental response data. Variability in response data that cannot be explained by information (here variability) in perturbation data might come from real ontic (*i.e.,* quantum[4]) randomness that manifests at relevant scales (Albrecht and Phillips, 2014), or simply from incomplete information about the actual perturbations to our system(s). Related to explanantia, uncertainty comes from the fact that we can never expect to perfectly characterize the dynamical processes in either the system under study, or our measurement devices. In context of any particular experiment, we might call the former type of uncertainty *aleatory* and the latter *epistemic*, where aleatory uncertainty refers to intrinsic randomness (here intrinsic relative to the experiment) and epistemic uncertainty is due to lack of perfect knowledge (here relative to the hypotheses, which are the explanatory objects) (Kiureghian and Ditlevsen, 2009).

Note that the probability distributions around all of the process components (*i.e.,* $\hbar_*$) in **Figure 1** represent epistemic uncertainty, and it is by marginalizing over these distributions that we are able to obtain probability distributions over phenomenological components, and finally over response data $z_y$ conditional on control data $z_u$ and $z_\theta$ as:

$$\mathcal{M}(z_y|z_u, z_\theta) = \int \int \hbar_y(z_y|y) \int \hbar_s(y|u,\theta) \int \hbar_u(u|z_u) \int \hbar_\theta(\theta|z_\theta) d\hbar_y d\hbar_s d\hbar_u d\hbar_\theta \mu_{\theta,u,y}. \qquad [1]$$

These predictions about actual data like $p(z_y|z_\theta, z_u)$ are the only things that a scientist can actually test. Further, aleatory uncertainty will manifest in $p(z_y|z_\theta, z_u)$ only if the various explanantia (*i.e.,* the various $\hbar_u$, $\hbar_\theta$, $\hbar_s$, and $\hbar_y$) are themselves probabilistic, and this is fundamentally different from recognizing epistemic uncertainty in the choice of any specific set of explanantia since we may have a correct or incorrect probability distribution conditional on experimental perturbations.

### 2.3. The Problem with Uncertainty

There are apparently two ways to test models based on their predictions: either by inter-comparing several models or by treating each hypothetical model individually. In reality, these approaches are not different, because any family of models that we might want to compare is itself the model that we end up testing. However, to facilitate a discussion of the two main approaches to testing models that currently exist, the following subsections draw a distinction between *Bayesian* (*e.g.,* Howson and Urbach, 1989) and *falsificationist* (*e.g.,* Mayo, 1996) types of inference. This distinction is described by Gelman and Shalizi (2013), who recognize that the statistics community sometimes incorrectly refers to the former as *inductive* and the latter as *hypothetico-deductive*.

The latter authors argue that there are certain inconsistencies between these two normative accounts vs. the way that they are actually applied in practice. Their poignant and illuminating treatment of this discrepancy between theory and practice nevertheless does not solve any underlying problem – instead of building a coherent normative theory of inference, they offer a largely ad hoc account of using two imperfect methods in tandem to help mitigate the deficiencies of each. Our argument in the following is that these differences between theory and practice betray deep underlying logical inconsistencies in both of the existing normative theories (Bayesian and falsificationism), and that such logical inconsistencies can be attacked directly by using a more complete system of logic.

#### 2.3.1. The Problem with Uncertainty for Bayesian Inference

Bayesian inference compares some number of competing models by first assigning over this family a probability mass function or probability density function, and then conditioning that probability distribution on observation data. That is, it is suggested that we may recognize competing versions of things like $\hbar_u$, $\hbar_\theta$, $\hbar_s$, and $\hbar_y$ as somehow representative of epistemic uncertainty, and that by modifying our doxastic states distributed across these alternatives based on reconciliation with observation data will lead us to reliable inferences.

---

[4] Quantum mechanics is the only known potential source of ontic randomness – all other randomness is epistemic (Jaynes, 2003; ch10).

The problem with this approach is that it is impossible to guarantee that any probabilistic inferences will be correct, or even consistent in any meaningful sense. Bayesian posteriors converge almost certainly to a true model, but this can only occur if a true model is assigned finite probability by the prior. This would require that the various epistemic distributions in **Figure 1** over competing explanantia like $\hbar_u$, $\hbar_\theta$, $\hbar_s$, and $\hbar_y$ be constructed such that they each place finite probability on at least one true process-representation. This alone is obviously unrealistic in all conceivable situations, however the problem is further complicated by the fact that any "true" explanantia must correctly represent aleatory uncertainty *for each experiment that we put to it*. So not only must we have a "true" process model, we also require a "true" statistical model of information missing from our experimental data *for each and every set of experiments*.

When these conditions are not met, probabilistic inference over families of models is not guaranteed to converge (Berk, 1966), and if inference does converge, may not converge to models that give the best predictions (Oreskes and Belitz, 2001, Grünwald and Langford, 2007, Müller, 2013). Even if we ignore Hume's (1748) problem and suppose that there does exist some space-time stationarity in the physics of the universe, so that that true models remain true, this does not mean that best *approximations* over past data will be best, or even reasonable, *approximations* over future data. For that, the error properties of our models would also have to be correctly characterized, which is indistinguishable from having a perfect model.

This is exactly the problem of verisimilitude (Popper, 2014), which we will discuss further in **Section 4**. There is no sense in which any individual proposition may be said to be more or less "truthlike", only that families of propositions (here models) contain as derivatives more or fewer true propositions. Popper proposed his theory of verisimilitude precisely to account for the difference between predictive adequacy and the "truthlikeness" of a model. The theory fails exactly because it only accounts for the nature of predictive statements that may be derived from a model.

In simplest terms, the problem is that in a Bayesian framework we have no way to understand the relationship between our family of models and the true nature of either phenomenology or process in the system – we just search for the best among alternatives conditional on data. Gelman and Shalizi (2013) point out that the way scientists currently deal with this is by using a falsificationist type approach in conjunction with a Bayesian approach; to state it loosely, falsification is used for actual model testing and Bayesian methods are used for calibration. The falsificationist approach is discussed in the next subsection (**Section 2.3.2**).

This issue causes real problems during inference. In a complex modeling chain, where epistemic uncertainty exists about a number of interacting model components, any misspecification of the distribution over one component will unavoidably bias probabilities placed on other components. An example of this was given by Beven (2008), and that example is expanded in **Appendix A** of this article. In these examples, errors in the specification of measurement models like $\hbar_u$ and $\hbar_y$ result in errors in the probabilities assigned to process descriptions $\hbar_s$.

Because we cannot ever guarantee that any of our epistemic distributions meet criteria for reliable convergence to true models, we have at least the potential for inconsistency in any Bayesian inference problem. Given this, it is impossible to assign anything like *reliable* posterior probabilities over any particular family of hypothesis. The best that we can do is assign probabilities that are *conditional on the other imperfect model components*, which are themselves necessarily distributional and almost certainly misspecified. This is exactly the confirmation holism aspect of the Duhem-Quine thesis as it manifests in the context of Bayesian inference.

To summarize the problem, it is that we are asking about truth-values of families of models that we know are all approximations. There is a fundamental logical contradiction inherent to the Bayesian philosophy itself. Remember that probabilities are doxastic measures associated with the truth-value of a given set of propositions, and so this logical error exists in any type of experiment where we attempt to assign probabilities to models. There is a contradiction between (i) the fact that we generally don't have access to anything like a "true" model, and (ii) the assignment of finite probabilities to models that we know with near certainty are incorrect. In fact, probabilistic inference methods assign doxastic states conditional on all information input to the problem, and given that any family of models is itself a model, there is no way to connect inferences over families of models with real epistemic uncertainty.

To reiterate what we said in the introduction, we are not advocating a strict anti-realist perspective on models – models may be truth-apt (Cartwright, 1983) in a meaningful way – it's just that since all models that we might actually construct are false (*i.e.,* models invariably represent reality in some simplified or incomplete manner), it might be worthwhile to recognize this formally in our method of science.

### 2.3.2. The Problem with Uncertainty for Statistical Hypothesis Testing

Consider next the case where we treat each model individually. Popper's (1959) motivation in developing a falsificationist perspective was the asymmetry between asserting truth vs. assessing falsehood. In particular, given Hume's problem (*i.e.,* that we can never guarantee that any generalization of data will apply outside of the domain of observation), all we can do under the classical logic described in **Section 2.1** is to reject models that disagree with those data that we actually have in hand. The general strategy of naïve falsification might therefore be thought of as rejecting models that conflict with data, and while this would indeed solve both the practical and logical inconsistency problems discussed in **Section 2.3.1** – since it doesn't require that we express any degree of belief in the *truth* of any model – it does require that we have a logically consistent way to reject models.

However, if we took this purely deductive approach and restricted ourselves to falsifying models directly via the modus tollens[5], then essentially all models would be immediately falsified given only a small number of data. This is true *even if we tested a perfect model*, since the best a modeler can ever hope to do in the presence of aleatory uncertainty is to build a model that predicts with accuracy and precision warranted by the (partial) information content of the experimental perturbation data. Somewhat ironically, while confirmation holism tells us that no *hypotheses* can be rejected, it is also the case that all *models* will be rejected given any data.

This following point is crucial: Bayesian methods cannot deal with epistemic uncertainty, in that they cannot tell us about anything about differences between model families and truth, and deductive methods cannot deal with aleatory uncertainty. This means (as Popper was well aware) that we must therefore apply some probabilistic or possibilistic version of falsification, and the result is that in general we are unable to strictly falsify any model. There is no longer any a priori criteria for rejecting models, and as Neyman (1957) put it, rejecting a model is an act of will rather than an act of rationality because the choice of rejection criteria is necessarily ad hoc. The problem is, however, worse than ad hoc rejection criteria – to have any criterion at all, ad hoc or otherwise, we must have some calculitic logic that allows us to develop some measure that may be associated with a given model and used as the quantitate basis for rejection. But now we are back where we started – we can't strictly falsify models, so instead we must associate with them some doxastic measure in the presence of incomplete experimental information.

That being said, it is clear that some models *are* useful, and so one might suggest that it matters little if the pedant (us) calls them false. We propose, however, that the practice of assigning degrees of belief to the Boolean truth-values associated with hypotheses or models that we know are incorrect is a primitive logical error that gives rise to essentially all of the practical problems we currently face related to evaluating the accuracy, precision, and reliability of complex systems models. Our point is that we should not use any method of inference that either implicitly or explicitly asks about the truth-value of any explanantia, and instead we wish to employ a quantitative method that recognize that treats this non-realism strictly. *We should be measuring degrees of isomorphism, not degrees of belief.*

In other words, our objective is a method of science that a) does not require (but may allow for) comparing competing explanantia, and b) does not assign degrees of probability or confidence to models. Certainly, we want this method to be fully coherent with probability theory, given Cox' theorem and that we will use the classical logic: we do not *necessarily* wish to deny non-contradiction, although we certainly leave open that possibility (*e.g.,* Kosko, 1990).

---

[5]The *modus tollens* says if proposition $P$ implies proposition $Q$ and $Q$ is false, then $P$ is false: $(P \rightarrow Q) \wedge \sim Q \rightarrow \sim P$.

3. **The Solution: An Information-Based Science**

It is helpful to first notice is that there is a significant component of empirical logic that is missing from probability theory. Let's see this by example. Suppose that we plan to conduct a series of experiments – to be concrete we will roll two dice. Before observing the outcome of the roll, our epistemic framework behaves multiplicatively: there $6 \times 6 = 36$ possible outcomes. After observing the outcome, our epistemic framework behaves additively: we now require $1 + 1 = 2$ pieces of information (the actual outcomes of each experiment) to capture everything we now know about the state of the dice. In general, doxastic states – the measures of belief we place on all explicitly stated possibilities – behave according to a product rule before observing experimental outcomes and according to a sum rule afterwards. Knuth (2005) showed that this logarithmic collapse of doxastic states – from multiplicative to additive behavior over multiple experiments – is unique and general in the context of a distributive and unitary calculitic logic like probability theory. Given that this logarithmic change in behavior effected by observation is not actually a part of probability theory, it is difficult to see why probability theory should be a sufficient logic for conducting any type of empirical investigation.

Notice that this argument for the incompleteness of probability theory is different than some others. For example, Pearl (2001) claimed that *"the bulk of human knowledge is organized around causal, not probabilistic relationships"* and that probability theory lacks causality. Pearl wishes to require that epistemic theory be designed to accommodate a particular a priori ontological view of the world (that physical laws are causal). Of course, we might counter-argue, as Hume did, that indeed all we may know are propensities. Whereas Pearl requires scientific epistemology to accommodate his a priori ontology, instead our argument is that no matter how we construct theories and models, probabilistic logic itself is incomplete because it doesn't recognize a fundamental characteristic of belief itself – how the behavior of related belief states change during experiment.

To state this symbolically, two experiments result in data $z_1$ and $z_2$ and our epistemic state changes due to observation as:

$$p(z_1, z_2) = p(z_2|z_1)p(z_1) \rightarrow h(z_2|z_1) + h(z_2) = h(z_1, z_2), \qquad [2]$$

where $h(.)$ is some measure of the information necessary to describe the outcome of our experiment. Given that this is a fundamental and necessary consequence of the epistemic environment described in **Section 2.1**, the question is how to account for it in practice.

3.1. **An Information-Based Scientific Method**

Let's start by asking what is probably the most straightforward question a scientist can ask in the context of any particular experiment: *Is there any information in my experimental data that could be used to improve my model?* If the answer is *yes* then we reject the model in a strict sense because we know that this model has the potential to be improved given only currently available experimental data. If the answer is *no* then this means that we have not discovered any potential to mitigate deficiencies in the model.

**Figure 2** illustrates a scientific method structured around this question. Again we have some experimental perturbation data and some experimental response data, and again these are not immediately treated as representative of anything in particular (data are ontic states of affairs rather than propositions about states of affairs). The perturbation and response data are actually related according to the physics of whatever system we are trying to model, and thus variability in perturbation data at least partially informs variability in response data. Since the objective of science is apparently to understand process relationships (Davies, 1998), ideally we would measure the information about the *relationship* between experimental perturbations and responses that is actually contained in that data, and then compare this to the information about this same relationship as it is described by our model. We typically can't do this directly, and so we do what is called *science*: we use tests of model *predictions* to reflect on the viability of model *explanantia*. We consider the act of using predictions to test hypotheses an essential component of any definition of science.

In this case, we want to compare information *about process relationships* contained in data vs. models, but will instead measure information *about experimental response data* that is available purely from relationships mined from data vs. information about experimental response data available from our hypothetical model that we want to test. Specifically, we will measure the information about experimental response data ($z_y$) contained in experimental perturbation data as some quantity $I(z_y; z_u)$ (for simplicity we will only notate $z_u$, but this could include any experimental perturbation data, including $z_\theta$), and the information about experimental response data contained in predictions made by model $m: z_u \to \widehat{z_y}$, that maps $z_u$ onto a prediction $\widehat{z_y}$, as some quantity $I(z_y; \widehat{z_y})$. If we use a self-equitable measure to quantify information (Kinney and Atwal, 2014) then information from the model is always bounded by the actual information content of the input data by the Data Processing Inequality (Gong et al., 2013):

$$I(z_y; z_u) \geq I(z_y; \widehat{z_y}). \qquad [3]$$

Supposing that we could estimate these two quantities: $I(z_y; z_u)$ and $I(z_y; \widehat{z_y})$, then the sign of their difference is sufficient to know whether information from the model is less than information from data. Again, a "yes" answer indicates potential to improve the model without collecting any new data.

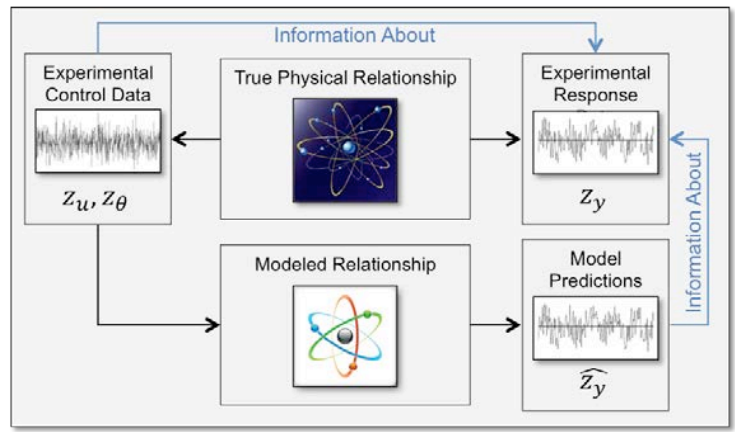

*Figure 2*: *An information-based science experiment. Experimental perturbation data partially informs experimental response data according to the physics of the real system, including all measurement devices. The model attempts to emulate this relationship, and, by the data processing inequality, model outputs contain no <u>more</u> information about experimental response data than is contained in experimental perturbation data. If the model provides less information than we are able to extract empirically from experimental data perturbation/response pairs, then the model is improvable given only currently available data.*

The challenge is to measure the information content of experimental perturbation data: $I(z_y; z_u)$, which requires that we know exactly the joint distribution between control/response data pairs. We obviously do not know this relationship exactly because if we did then we would not have need to test any further models. Instead we return to the actual desideratum, which is to test relationships rather than predictions, and ask simply whether the model provides as much information about the *relationship between experimental perturbation and response data* as we are actually able to extract by building a relationship purely from that data itself (Nearing and Gupta, 2015). The following explains.

Although we can't estimate $I(z_y; z_u)$ directly, we can bound it conservatively. For example, the current authors have approached this using nonparametric regression (*e.g.,* Nearing et al., 2016a, Nearing and Gupta, 2015). Given an arbitrary empirically-derived mapping $r: z_u \to z_y$, then by another application of the data processing inequality we have $I(z_u; z_y) \geq I(z_y; r(z_u))$ such that information missing from model $m$ is bounded:

$$\mathcal{E} = I(z_y; z_u) - I(z_y; \widehat{z_y}); \qquad [4.1]$$
$$\hat{\mathcal{E}} = I(z_y; r(z_u)) - I(z_y; \widehat{z_y}); \qquad [4.2]$$

$$\varepsilon \geq \hat{\varepsilon}. \qquad [4.3]$$

If $r$ has known convergence properties over large function classes (*e.g.,* Hornik, 1991), then in principle and as long as metric $I$ itself admits a convergent estimator, we have a bounded and convergent estimate of the information content of experimental data. In practice, of course, we always must approximate by either optimizing or integrating over regression hyperparameters, and so our results are not actually convergent, but unlike purely probabilistic methods we have a *logically consistent theory* that we can at least approximate – and we are not *in principle* precluded from accomplishing our objectives as we were when conducting an experiment like in **Figure 1**. Not only that, this method is robust (*i.e.,* bounded) to approximation in all cases except overfitting, which can be mitigated at least empirically by partitioning the experimental data into calibration and evaluation records when calculating $I(z_y; r(z_u))$.

It's worth pointing out that we really want to have a theoretical convergence proof for whatever strategy we use to estimate $I(z_y; z_u)$ from data. Ideally this would include theoretical convergence rates as well as an expression for the asymptotic distribution over the statistical estimator $I(z_y; r(z_u))$ in the limit of an increasing number of data, and perhaps even a series expansion that yields ordered bounds. However, such theoretical demonstration may or may not be possible for any particular estimator, and we will leave this as an open challenge. However, again, the point is that the basic theory is consistent and approximable even without a demonstration of asymptotic behavior.

To summarize, a meaningful model evaluation experiment might seek to establish whether the model contains (at least) as much information about the relationship between experimental perturbation/response data as do the data themselves. Because we cannot directly measure information about relationships, we instead measure information about experimental response data that is contained in experimental perturbation data, and compare that with information contained in model predictions. Of course, we cannot measure information contained in experimental perturbation data without a model of the relationships underlying that data, and so our objective is to develop an empirical relationship between experimental perturbation data and experimental response data that is completely theory-free, or at least based solely on logical theory rather than ontological theory, and which seeks to extract all possible information about the perturbation/response relationship as possible from experimental data. We then measure the information about response data provided by whatever relationship we are able to extract purely from data vs. from our hypothetical model, and this conservatively bounds the information missing from the hypothetical model.

### 3.2. Measures of Information for Hypothesis Testing

In **Section 3.1** we only required that information be quantified by a self-equitable metric so that we can employ the data processing inequality (Kinney and Atwal, 2014). This is, however, insufficient for logical consistency since we must account for Knuth's (2005) description of the logarithmic relationship between information and probability (**Equation [2]**). This constraint is helpful because it reduces apparent freedom of choice about what information metric to use.

Under probability theory, everything that we know about any potential data $z_y$ is described by some probability distribution $p(z_y)$ – this distribution is an expression of available information before applying any explanatory or predictive hypotheses or conditioning on any exogenous data, like model inputs. Here we don't care about whether a true outcome is assigned finite probability, we only care that $p(z_y)$ represents the distribution of our actual experimental response data $z_y$ in absence of any other data or any hypotheses – this is a frequency distribution. Next we obtain some data, say $z_u$, and appropriately condition $p(z_y)$ to obtain $p(z_y|z_u)$. The resulting reduction in new information that we *would need* to identify any particular occurrence of $z_y$ is, by **Equation [2]**, the difference $h(z_y) - h(z_y|z_u)$. We can compare this difference with the reduction in new information needed to fully determine $z_y$ if, instead of conditioning on $z_u$, we were to condition on model predictions $\widehat{z_y}$: $h(z_y) - h(z_y|\widehat{z_y})$. The information that is available in model inputs that is missing from model predictions is $h(z_y|z_u) -$ $h(z_y|\widehat{z_y})$ and, according to the data processing inequality, this quantity is always non-negative as long as the conditional and shared information measures behave additively as: $I(z_y; z_u) = h(z_y) - h(z_y|z_u)$. This is precisely the relationship that is satisfied by the standard Shannon-type entropy and mutual information metrics (Cover and Thomas, 1991).

It is useful to develop an epistemological perspective on the uniqueness of Shannon's entropy and mutual information measures, but first let's define them. Entropy is the expected amount of information about a random variable that can be gained by observing that variable, and is defined as:

$$h(z_y) = \int p(z_y = \zeta) \ln\left(p(z_y = \zeta)^{-1}\right) d\zeta, \qquad [5.1]$$

So, entropy is (in some sense) a measure of the variability in a probability distribution over a random variable[6]. Following from this, mutual information is the expected reduction in entropy of one random variable that is caused by conditioning on a correlated variable; *e.g.*:

$$I(z_y; z_u) = h(z_y) - h(z_y|z_u), \qquad [5.2]$$

$$I(z_y; z_u) = \int \int p(z_y = \zeta, z_u = v) \ln\left(\frac{p(z_y = \zeta|z_u = v)}{p(z_y = \zeta)}\right) d\zeta dv. \qquad [5.3]$$

So, mutual information is the expected amount of information about one variable that is contained in a realization of a correlated variable, and so $I(z_y; z_u)$ specifically measures the expected reduction in the amount of entropy that would be achieved by directly observing $z_y$ depending on whether we had first conditioned on $z_u$. The quantity $I(z_y; \widehat{z_y})$ is similar, except here we are concerned with model predictions such that the model predictions are independent of experimental response data conditional on experimental perturbation data.

Notice that $I(.)$ is a special case of integration over a transformed ratio of a conditional distribution to its marginal:

$$I(z_y; z_u) = E\left[f\left(\frac{p(z_y|z_u)}{p(z_y)}\right)\right]. \qquad [6]$$

There are three important component attributes of this statistic. The first is the probability ratio itself, which is simply the expression of the relationship between marginal and conditional knowledge about $z_y$ given $z_u$. The second important aspect is the integration itself. In general the probability ratio is high dimensional (infinite-dimensional in the case of continuous random variables), and the integration collapses such distributions to a metric. The integration is purely for tractability, and thus we may want to integrate several different statistics to understand different aspects of the probability ratio. The third important aspect of this statistic is the log-transform. While the conversion from a distribution to a metric is general as long as we allow for any transform of the probability ratio inside of our integration, we will specifically obtain a self-equitable metric as long as our transform is convex (Csiszár, 1972). Here, however, we must use Shannon's logarithm, $f(u) = -u\ln(u)$, to be consistent with **Equation [2]**.

Further, it is important to stress that the above information metrics are not bridge principles. Bridge principles are tools used to join together different bodies of theory (Carnielli and Coniglio, 2007), and typically, model evaluation metrics are treated as epistemic bridges that relate the model's ontological predictions with appropriate epistemic consequences (*i.e.,* probability distributions). In contrast, the above information metrics derive uniquely from the same system of logic that supports probability theory in the first place. That is, if our epistemic theory admits probabilities, then we do not need any bridge principles to quantify model performance, we are forced to use information theory. Moreover, by using this theory, we get the scientific method outlined in **Figure 2**, which avoids essentially all of the pitfalls discussed in **Section 2**.

---

[6] We won't pretend that statistical entropy is in any way related to uncertainty, even though we have used that language in previous publications. The reason that this is incorrect is again that the distributions like $p(z_*)$ that are integrated to get quantities like $h(.)$ suffer from the usual degeneracies, and therefore have no systematic relationship with uncertainty. We simply drop the concept of uncertainty altogether – it is not useful.

It is interesting to notice that linearized epistemic bridges are very common (*e.g.,* Tian et al., 2016, Taylor, 2001, Gupta et al., 2009), and these are chosen apparently for mathematical convenience. What we have argued above is that there is a *real* (not arbitrary) linearity in our epistemological theory that is exploited to result in additive measures of model performance. This linearity does not result from convenience, but rather results necessarily as a fundamental property of distributive logic. While Nearing & Gupta (2015) showed that the standard linearized evaluation metrics (mean-squared error, Pearson product-moment correlation coefficient, additive mean bias) are special cases of the integration in **Equation [6]** that come from assigning specific integrating functions $f(.)$ and specific parametric forms of the conditional and marginal probabilities, the only integrating function that makes **Equation [6]** coherent with **Equation [2]** is Shannon's: $f(u) = -u \ln(u)$.

In particular, the metrics in **Equation [5]** are apparently unique under the following choices:

- First, the choice to use a two-valued logic that includes non-contradiction, so that propositions are either true or false but not both. An example of an alternative to this is Fuzzy Logic (Kosko, 1990).

- Second, the choice to represent our dynamic beliefs about various propositions using a scalar belief metric (Van Horn, 2003). Then by Cox' theorem we have probability theory, and by Knuth's theorem probabilities collapse logarithmically over repeated experiments.

- Third, the choice to work with metrics rather than distributions. The integrations that turn distributions into statistics results in Shannon's information theory.

This does not, however, mean that there are no choices left to make. The fact that we are quantifying information by taking an *expected value* (*i.e.,* an integration over the distributions) is important. Any statistical procedure requires some representative sample, and so we must integrate our statistics over some finite number of experimental trials – and the choice of integration domain matters. That is, our results will depend on which experimental data we consider. Ideally we would consider all existing experimental data that is in any way relevant to our hypotheses, but in practice this is obviously impossible. Again, however, the point is that we can make reliable (bounded) hypothetico-deductive inferences in the context of whatever experimental data we do decide to use. Moreover, we can actually use differences between integration domains to quantify process stationarity (*e.g.,* Nearing et al., in prep).

Another important choice is about which aspect(s) of our experimental data we want to investigate. There is apparently nothing that prevents us from manipulating data before calculating information statistics. For example, we may be concerned only with extreme events, and we may therefore assign phenomenological variable $y$ to represent some aspect of our definition of "extreme". Or we may care about time-series matching, in which case $y$ might represent some aspect(s) of a power spectrum. In all cases, our model will predict the particular aspect of the experimental data that we want to test, and we may also actually manipulate the actual experimental data (*e.g.,* calculate its power spectrum). Notice that manipulating experimental results does not actually constitute applying a data model or measurement model – it is simply physical manipulation of data via some information-processing device like a computer, and we will simply test the extent to which our holistic model accounts for whatever eventual data we put it up against.

The result of this is apparently a theory of hypothesis testing that solves two basic problems: 1) we can test models that do not inherently admit probabilistic predictions (thereby avoiding the degeneracy problem related to separating aleatory uncertainty), and 2) we do not require any ad hoc epistemic bridges or ad hoc falsification criteria. Our falsification criteria here are never perfect (obtained through high-dimensional nonparametric regression) but if calculated reliably, they always provide a lower bound on the available information, and therefore, at least in theory, we can test models without any chance of Type I error. A toy application of this theory is given in **Appendix B**. Note that this example extends the inconsistency example in **Appendix A**.

4. **Right Answers for the Right Reasons**

In **Section 3.1**, we made a distinction between testing models based on their ability to inform process relationships vs. their ability to make informative predictions. The risk in doing the latter is that we may build models that are informative over particular data sets for wrong reasons – either by chance or by *kludging*. Kludging happens when informative predictions are derived from model behavior that is not isomorphic with real system dynamics (Clark, 1987).

Of course, potential for kludging is not unique to the theory we've proposed. For example, Popper (2014) dealt with the fact that our best scientific theories and models are often strictly false by defining a concept of *verisimilitude*, or approximation of truth. According to Popper, one theory is a better approximation of truth than another if it entails more true (presumably phenomenological) statements. This fails to solve the problem, however, because testing the truth of entailed propositions is exactly the definition of testing a model's ability to make true predictions.

As far as we can tell, the problem of assessing truthlikeness or verisimilitude is intractable. Instead, we might approach the problem from a reductionist perspective. Under the assumption that there does exist regularity in the universe, and that it is the objective of a scientist to discover this regularity, then we not only want informative models, we want models that are reliably informative because they generate predictions using *processes that are isomorphic with whatever dynamic aspects of the system are actually process-stationary*. This is what the reductionist means when we say that we want right answers for the right reasons, and is the only feasible concept of reliability that the current authors can imagine.

So can we at least measure process isomorphism in physical systems? Again, we can't measure information about individual process relationships embedded in a model, but we can measure information transferred through individual process relationships. This is not unlike measuring information provided by model predictions except that here we don't have anything like experimental perturbation data, since in any complex model with many interacting process relationships the inputs to each process component are themselves modeled variables. We cannot simply dissect a complex systems model and run each component separately, because the model must include interactions between all of the various hypothetical process relationships. Although environmental models often do not contain explicit field equations, the governing conservation equations will generally include many directly or indirectly interacting effects.

So we will not ask whether particular process relationships embedded in a model individually produce informative predictions; instead we will treat models as networks of interacting processes. Perhaps those inclined to Pearl's perspective on causality might call these *causal networks*, and recognize that they cannot be evaluated in the same way as holistic models because of the lack of experimental perturbation data on the model side.

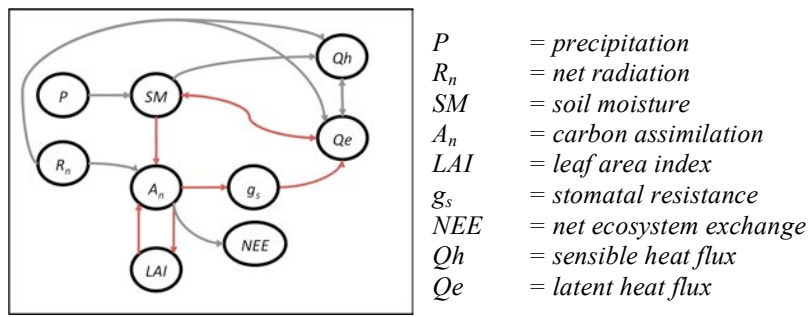

| | |
|---|---|
| P | = precipitation |
| $R_n$ | = net radiation |
| SM | = soil moisture |
| $A_n$ | = carbon assimilation |
| LAI | = leaf area index |
| $g_s$ | = stomatal resistance |
| NEE | = net ecosystem exchange |
| Qh | = sensible heat flux |
| Qe | = latent heat flux |

*Figure 3*: *A process network representing a typical ecohydrology model. Nodes represent phenomenological variables and edges represent components of explanantia.*

More generally, because we are using probability theory we will treat process networks as so-called Bayesian networks. An example of a Bayesian network of a typical ecohydrology model is given in **Figure 3**. Each node in this network represents a phenomenological variable with some spatiotemporal extent, and we represent variability (not uncertainty) in each node using probability distributions that are conditional on all variables upstream in the network. The edges in the network represent explanantia that manifest as probabilistic conditioning relationships. Again, these probabilities do not represent uncertainty but rather partial informativeness – the conditioning variable informs observed or modeled variability in the conditioned variable. Even if all equations in the model are deterministic each variable is still probabilistic conditional on only a subset of whatever other variables determine its value at any point during the simulation. Our objective is to quantify the influence that each variable has on all others in the real world, and then to see whether the model reliably simulates these partially informative relationships.

To quantify the influence that one variable, say $y_1$, has on another variable, say $y_2$, in a dynamic (time-evolving) Markovian (*i.e.,* local or causal) system, we would measure the expected effect of conditioning $y_2$ at time $t+s$ on the value of $y_1$ at time $t$ given that all of the variables in the model other than $y_1$ (notate these as $y_{\sim 1}$) also had particular values at time $t$. We therefore measure the influence of $y_1$ on $y_2$ in a Markov system over some time lag $s$ as:

$$I(y_2^{t+s}; y_1^t | y_{\sim 1}^t) = \int p(y_{\sim 1}^t) \int \int p(y_2^{t+s}, y_1^t | y_{\sim 1}^t) \ln \left( \frac{p(y_2^{t+s} | y_1^t, y_{\sim 1}^t)}{p\left((y_2^{t+s} | y_{\sim 1}^t)\right)} \right) dy_2^{t+s} \, dy_1^t \, d y_{\sim 1}^t. \qquad [7.1]$$

This metric typically has to be approximated due to a problem of dimensionality ($y_{\sim 1}$ is high-dimensional), and the most common approximation is called *transfer entropy* (Schreiber, 2000):

$$I(y_2^{t+s}; y_1^t | y_2^t) = \int p(y_2^t) \int \int p(y_2^{t+s}, y_1^t | y_2^t) \ln \left( \frac{p(y_2^{t+s} | y_1^t, y_2^t)}{p\left((y_2^{t+s} | y_2^t)\right)} \right) dy_2^{t+s} \, dy_1^t \, d y_2^t. \qquad [7.2]$$

Transfer entropy relies on a strong Markovian approximation whereby the time history of the whole model, $y_{\sim 1}^t$ is substituted by the time history of the target variable, $y_2^t$. The result is that this metric only ever requires integration over a 3-dimensional probability distribution, and is therefore generally feasible to estimate given a realistic number of data.

**Equations [7]** calculate the strength of directed interactions between each pair of phenomenological variables in our model at any spatiotemporal scale (if desired we can substitute the time indexes for spatial indexes). If we had complete observations of all phenomenological variables we could construct similar process networks from data and compare the strengths of interactions derived from data with the hypothetical strength of interactions from our model. The model may either over-estimate or under-estimate the informativeness of any one variable about any other, and both over-estimation and under-estimation in this case is undesirable.

Of course, we must deal with the very real problem that we will never be able to directly observe every phenomenological component of our modeled system, and therefore cannot expect to actually construct Bayesian networks or calculate all necessary transfer entropy metrics directly from observations. However, we can do this with whatever (partial) observations we do actually have available, and again the point is that we have a coherent and general theory that we can approximate in practice. Further, we propose that we can even take advantage of partial observations by using data assimilation-based system identification, which is the application of Bayes' theorem to condition the internal states of a model on information in whatever observations we do actually have available (*e.g.,* Bulygina and Gupta, 2011, Wilkinson et al., 2011). The idea being to measure how conditioning model phase-space trajectories change (Wikle and Berliner, 2007) conditional on information in observation data, and how this implies changes to process-level information transfers within the model.

A simple example of this theory is given in **Appendix C**, while Drewery et al. (in prep) give several more sophisticated examples. We will not go into further detail here except to say that this type of Bayesian system identification does not suffer from the problem of inconsistency under degeneracy, because we are simply asking how observations project onto the phase trajectories of whatever model we have proposed. We are not using data assimilation to search for a "true" phase trajectory or "true" Bayesian information transfer network, we are using data assimilation to ask how much our observations can inform changes to modeled phase trajectories and modeled process-level information transfer.

It's worth pointing out that this diagnostic approach (with or without data assimilation) does not tell us whether we are missing any important processes in the model. However, it does help us understand whether whatever processes we have hypothesized and modeled interact in ways that are apparently isomorphic with observed interactions. This is not a formal approach to gaining knowledge, only an approach to gaining diagnostic insight into the structure and functioning of complex models.

## 5. Discussion: Science Arbitration and Predicting with Limited Information

In the preceding sections we argued for a particular perspective on evaluating models for the purpose of scientific learning. However, another purpose of building science-informed models is to make predictions that contribute to societal management and decision strategies. This is called *science arbitration* (Pielke, 2007), and it might seem inevitable that successful and comprehensive model-based arbitration requires communicating predictive uncertainty to decision makers. We argue that this is a somewhat misleading prescription. Instead, what the scientist actually wants to do is to communicate to decision-makers predictions that represent the scientist's best available information. *Such predictions will necessarily be distributional, but will not be representative of uncertainty in any meaningful sense*. The scientist should also communicate all choices that were made during model development and evaluation – for example by using formal decision trees (Beven, 2016).

So how do we predict with our best available information? The first thing that we must do is to ensure that we do not pretend to have more information than we actually have. We should not over-constrain our models. The practice of producing deterministic predictions and then appending to these "uncertainty distributions" or "error distributions" is very strange indeed. This practice essentially boils down to over-estimating our available information in the first place and then adding more information about that overestimation. Instead, a better organizing question seems to be about how we might construct models that do not require that we pretend to have more information than we actually have from theories, hypotheses, and data.

Methodologically this perspective implies a very different approach to model development. In particular, we suspect that this perspective will lead to a strategy for constructing models of complex systems that does not use differential equations to represent conservation laws. Instead we expect that in the not-too-distant future the general approach to constructing models of conservative macro-scale dynamical systems will involve imposing conservation laws expressed as symmetry constraints, via Noether's theorem, on maximum entropy distributions.

The motivation for using maximum entropy distributions is that the basic project when developing a model under a calculitic epistemic theory is to construct a joint distribution between all modeled variables over the entire (perhaps spatiotemporal) domain of the simulation. In the case where we know absolutely nothing about the system, this distribution will be non-informative. Any theory, hypothesis, or data that we have or want to test will provide information about the behavior of the system, and should be used to constrain this joint distribution. In a discrete model (*e.g.,* like what a finite element or finite difference approach produces), this joint distribution would be a directed Bayesian network. The structure of the creative project in science is now to develop hypotheses about dynamic systems that act directly as epistemic constraints on Bayesian networks, instead of as flux or source/sink terms in governing or field equations.

## 6. Conclusion

To reduce the situation to a vague analogy, quantifying uncertainty is very much like trying to measure cold or dark – uncertainty is the absence of knowledge, and it seems natural that we might instead develop theories around the thing that we actually have access to: heat, light, or information. The constructive question – which we have started to answer here – is about how we accomplish our scientific inference and arbitration objectives in this context. After writing this paper, we struggle somewhat to understand why uncertainty quantification has enjoyed such a central place in both the philosophy and practice of science. The idea of assigning truth-values to hypotheses or models, and trying to assess our limited ability to do so seems to be a somewhat contrived or fantastical objective. It seems much more natural to try to understand and quantify what information we have available from data about our models. In general, information-based approaches to inference, arbitration, and communication seem much more coherent, straightforward, and intuitive than uncertainty-based approaches.

The proposal is to avoid associating doxastic states directly with things that we have any fundamental uncertainty about – like hypotheses, models, or future data. Instead, we recognize that probabilities are expressions of partial information rather than expression of partial uncertainty, and we use probabilities simply to quantify partially informative relationships between ontic states (*i.e.,* actually-existing hypothetical and experimental data).

It is essentially indisputable in the modern era that probability theory is a system of logic, and the theory we propose is absolutely built around this fact. However, our theory only ever *applies* probabilities to represent counting distributions. The reason for this is that under empiricism, we gain information by actual experience, and so when we account for the doxastic effects of experiment we are fundamentally counting. More specifically, we do assign doxastic measures, but only to phenomenological entities and not to theoretical entities, and then only based on empirically available information. By refusing to explicitly treat explanantia as truth-apt (although leaving open the possibility for any type of realism the scientist might prefer), we are taking an approach that is quite similar to the old frequentist interpretation of probability theory. This is perhaps not surprising since frequentism is a special case of the Cox-Jaynes interpretation of probabilities as expressions of rational doxastic consequences of partial information (Terenin and Draper, 2015).

Although Howson (2000) did not call on Cox or Jaynes directly, that same mature theory of probability theory as epistemic logic is what he was referring to when he said *"The 300-year-old programme for an inductive logic based on formal probability has arrived finally at maturity ... now, for the first time in its long history, it can display its own explanatory credentials as an authentic species of logic, kindred to deductive logic."* The development of this theory is almost certainly one of humanity's greatest achievements, however it is not a complete description of empirical inquiry without Knuth's (2005) demonstration of the relationship between the logic of questions (probability theory) and the logic of answers (information theory). Only now, in the beginning of the 21$^{st}$ century, are we entering an era where we have access to what seems to be a complete logic of inquiry, and although the problem of deriving non-ad hoc hypothesis tests is quite old, it does seem that the time is ripe for progress.

To restate our central claim, the only question that a scientist can coherently ask and reliably answer is: *"Does my model capture all of the information that is present in my experimental data?"*

**Appendix A: An Example of Bayesian Inconsistency**

The purpose of this toy example is to show that probabilistic inference is inconsistent under degeneracy. In particular, incorrect phenomenological distributions result in not only incorrect, but actually contradictory (under different assumptions), inferences over ensembles. We draw inspiration for this example from Beven et al. (2008), who showed that incorrect likelihood functions result in incorrect parameter estimates; their likelihood function is effectively an approximation of the result of integrating **Equation [1]**, and if we get this approximation wrong (in their example by using an incorrect phenomenological distribution over *y* appended to each parameterized model), then we cannot expect to obtain either correct or consistent parameter inferences. This is exactly an example of the Bayesian reliability problem that we are discussing here, and we demonstrate this same effect on assigning probabilities to competing models after marginalizing over parameters.

Later, in **Appendix B**, we extrapolate this example to demonstrate that information theory allows us to obtain reliable measures of information missing from experimental data and from model predictions, and to do this we use a synthetic experiment where the correct answer is known exactly. The synthetic inference problem was constructed using daily precipitation (in [mm]) and daily potential evaporation (in [mm]) data from the Leaf River catchment in Mississippi, USA to simulate a 1000-day streamflow record (also in mm) with the HyMod (Boyle, 2000) model that included three slow-flow tanks. The output time series from this simulation was taken as synthetic 'truth' data, and the record was split into a 500-day warm-up period and a 500-day observation period. The model outflow fraction parameters and were sampled uniformly over [0,1]; the soil moisture storage parameter was sampled uniformly over [0-1000] mm, and the infiltration exponent uniformly over [0,10].

The ensemble $\mathcal{M}$ consisted of two competing model structures: a three-bucket Nash cascade and the abc-hydrology model, both run at a daily timestep. In both models, the parameters are outflow ratios sampled uniformly over [0,1]; 500 parameter sets were sampled for each model. Additionally 500 different time series of daily precipitation were sampled from each of three different measurement distributions – all stationary iid

Gaussian with standard deviations of $\sigma_u = 0.01, \sigma_u = 0.1$, and $\sigma_u = 0.5$. We then calculated probabilities for each of the 500×500 = 250,000 model simulations from each of three input distributions using each of three different measurement distributions over streamflow with the same standard deviations as before: $\sigma_y = 0.01, \sigma_y = 0.1$, and $\sigma_y = 0.5$. In total there were nine experiments each consisting of 250,000 runs of each of two models.

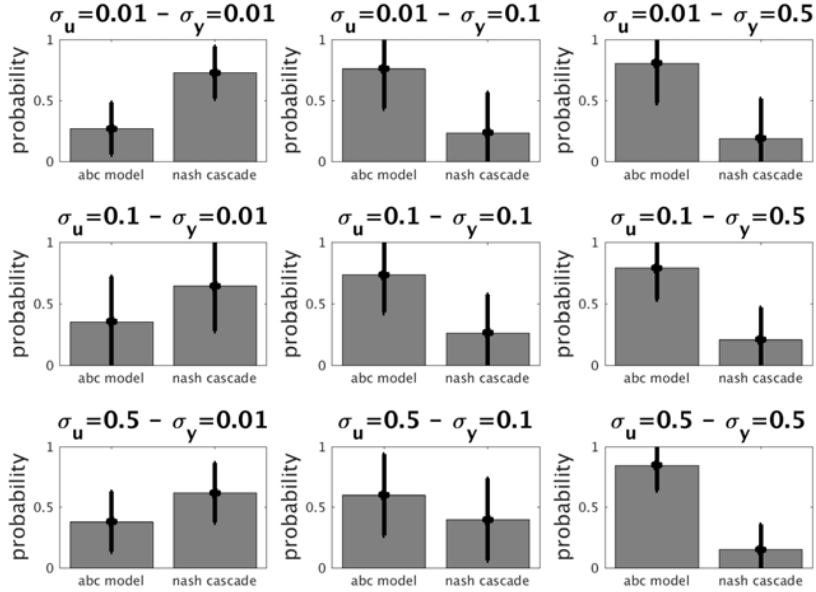

*Figure A.1*: Inconsistent results in a synthetic multi-model inverse problem. Model probabilities depend on the choice of phenomenological distributions over precipitation and streamflow. Error bars represent one standard deviation over bootstrapped samples of the observations.

Probabilities associated with each of the two competing models were calculated by bootstrapping daily observations from the observation period and then marginalizing over all 250,000 models parameter/boundary condition combinations in each of the nine experiments. The relative probabilities placed over the two-model ensemble are presented in **Figure A.1**. Ten bootstrapped observation samples were used to calculate the means and standard deviations that are plotted – bootstrapping was done to avoid undue influence of any individual observation, since probabilities are multiplicative over observations and we necessarily used a finite data record.

The take-away from **Figure A.1** is that we get completely different results in even the relative rankings of the two models depending on the choice of measurement distributions. In some cases the ABC model is assigned greater than 50% probability on average (across bootstrap samples) and in others the Nash cascade is assigned greater than 50% probability. There is no consistency. More generally, we will never in practice find ourselves in a situation where our models only have a single uncertain component, and for any situation where our modeling chain has interacting degeneracies in uncertainty distributions, inference results will always at least have the potential to be inconsistent.

**Appendix B: Consistency of Inferences under Information Theory**

Here we re-examine the synthetic example given in **Appendix A** from the information-based perspective outlined in **Section 3**. This is an example of what the information-based method looks like as it approximates correct answers to the inference problem under uncertainty.

As in **Appendix A**, daily precipitation [mm] and potential evaporation [mm] were used to create synthetic 'true' streamflow data using HyMod ($z_y$), and synthetic precipitation forcings were then generated by sampling perturbations from iid Gaussian distributions to generate several different forcing data sets ($z_u$). This time instead of sampling from three forcing distributions, we sampled from several with increasing standard deviations, in order to provide more resolution in the effect of limited information content of forcing data. Just as in **Appendix A**, several 3-bucket Nash cascades and abc-hydrology models were used to simulate daily streamflow over a 10,000-day period using these perturbed forcings

True measures of information missing from each set of synthetic precipitation data was measured by running the perturbed forcings through the real system (HyMod) and calculating the entropy of the synthetic observations conditional on each of the resulting streamflow series. These quantities are notated as $H(z_y|z_u)$. True measures of the information missing from each model were calculated by running each synthetic precipitation data series through one of our competing models – either a three-bucket Nash cascades or an abc-hydrology model, each with different parameters – to generate model predictions $\widehat{z_y}$. Parameter samples were the same as those used in **Appendix A**. The difference $H(z_y|z_u)$ and the resulting model-conditional entropies $H(z_y|\widehat{z_y})$ quantify the entropy due to model error according to **Equation [4.2]**. As in **Appendix A**, there were two model structures and 500 parameter samples.

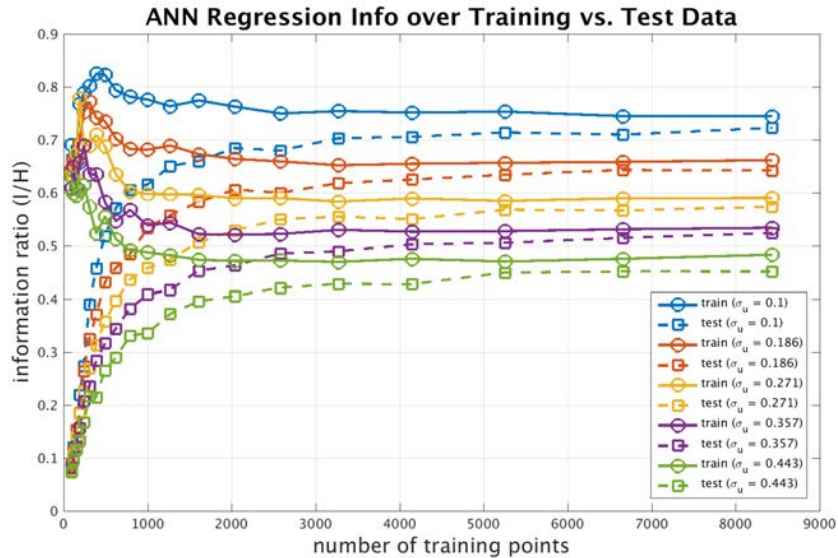

*Figure B.1*: *Convergence of mutual information statistics during calibration and testing data records after training neural networks to estimate $I(z_y; f(z_u))$.*

Again, the objective of this example is to demonstrate that we can accurately estimate information missing from the model using **Equation [4.1]**. The quantities $I(z_y; f(z_u))$ were calculated using single-layer feed-forward neural networks to regress each set of (lagged) precipitation data onto streamflow observation data. We used a ninety-day lag period, so that there were ninety precipitation inputs used to predict each streamflow output. It is important that our regression models are not overfit when used to bound $I(z_y; z_u) \geq I(z_y; f(z_u))$. To assess this, neural networks were trained on an increasing fraction of the total data record and the trained networks were used to predict both in-sample and out-of-sample data points. The objective is for the in-sample and out-of-sample information statistics to converge, which is what we see in **Figure B.1**. In particular, the in-sample and out-of-sample statistics converged to within 5% of each other with about 5000 training data.

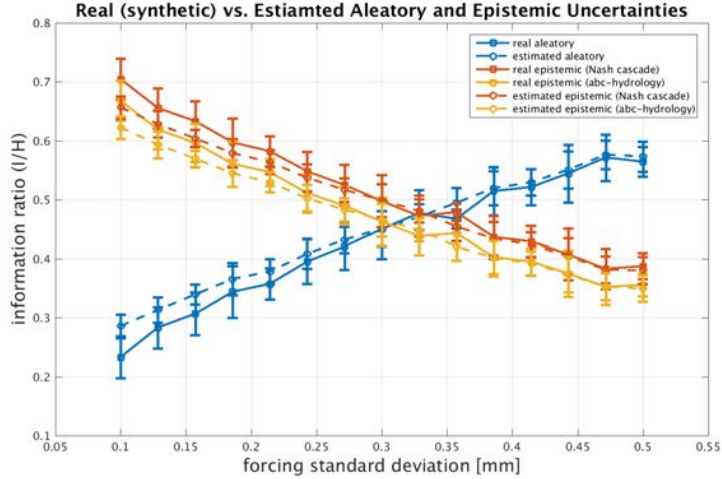

*Figure B.2*: *Real vs. estimated information missing from (i) experimental perturbation data and (ii) model predictions.*

Next, the out-of-sample predictions made by these trained networks were used to calculate $I\left(z_y; f(z_u)\right)$ statistics, which bound the $I(z_y; z_u)$ statistics that we really want. The mutual information between Nash cascade predictions and each different input data set, $I(z_y; \widehat{z_y})$, were then used in **Equation [4.1]** to bound information missing from the model according to $\hat{\mathcal{E}} \leq \mathcal{E}$. **Figure B.2** compares 'real' vs. estimated missing information (from experimental data and model predictions) as a function of the standard deviation of forcing perturbations. Error bars represent one standard deviation over 30 repeated experiments using different Nash cascade parameterizations (again parameters were sampled uniformly over [0,1]). Unlike in **Appendix A**, we treat each model individually and there is no need to marginalize over parameters (although this could de done if we wanted to test a model that included a parameter distribution). Information missing from the model predictions is always underestimated, as expected, but in this case the under-estimation is generally less than 5% relative error.

**Appendix C: An Example of Process-Level Diagnostics**

This appendix includes a demonstration of the verisimilitude theory outlined in **Section 4** applied to a simple rainfall-runoff model like what were used in the previous two appendices. In this case we used the HyMod model with real-world precipitation and streamflow data from the same Leaf River catchment. Since we are interested to understand somehow the realism of internal mechanics of this model, we need to have some understanding of the model itself. A conceptual illustration of the HyMod rainfall-runoff simulator is in **Figure C.1** - in this model a watershed is conceptualized as a set of water mass stores with linear outflow rates.

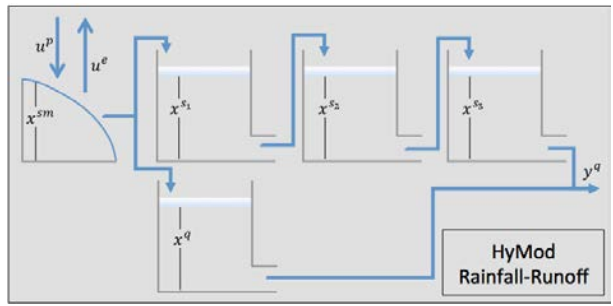

*Figure C.1*: *A conceptual diagram of the HyMod rainfall-runoff simulator. Markov state variables are labeled $x_*$, the precipitation boundary condition is $u^p$, the potential evaporation boundary condition is $u^e$, and simulated streamflow is $y^q$. Model parameters (not notated) control the height of a nonlinear soil moisture storage tank, the outflow ratios from each storage bucket, and the fractional partitioning of soil moisture into surface runoff ($x^q$) and subsurface runoff ($x^s$).*

**Figure C.2** shows what this same model looks like when conceptualized as a Bayesian network. The same phenomenological variables are labeled, and the network connections illustrate the paths of information flow during one timestep of numerical integration. The only observed variable is streamflow, and this is observed at the integration timestep (daily). Our objective is to use information from that single observed time series to learn about overall process-level deficiencies in the model.

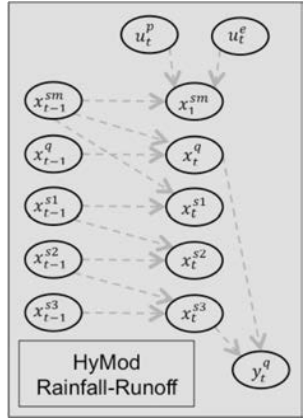

*Figure C.2: A Bayesian network illustration of the same model as in **Figure C.1**.*

Predictive accuracy (as quantified using a mean-squared error) of all of these procedures (calibration, data assimilation, and system identification) are illustrated in **Figure C.3**. It is, of course, entirely inappropriate to use a squared error metric in any way during the process of conducting science (see **Section 3.2**) but for illustrative purposes this is sufficient and sufficiently simple to make the point. The point is that calibration helps improve model performance both during calibration and evaluation (forecasting) periods. Data assimilation, which only updates initial states for prediction, only improved model accuracy during the calibration period where we had observations available to assimilate. System identification, on the other hand assimilated observations in the same way as data assimilation, however it encoded the information from those observations into the structure of the Bayesian network, which facilitated better predictions not only during the calibration period, but also during the evaluation period.

More to the point of what we want to do here, we can measure the process-level information flows within the model before vs. after system identification to understand what assimilated information from our observation data has to tell us about deficiencies in the model structure. We apply **Equation [7.2]** to calculate information flows along each edge of the Bayesian network illustrated in **Figure C.2** both before and after system identification. The differences between the values of the prior and posterior transfer metrics are illustrated in **Figure C.4**. This figure shows differences in information transfers between pairs of modeled variables that were effected by assimilating streamflow observations.

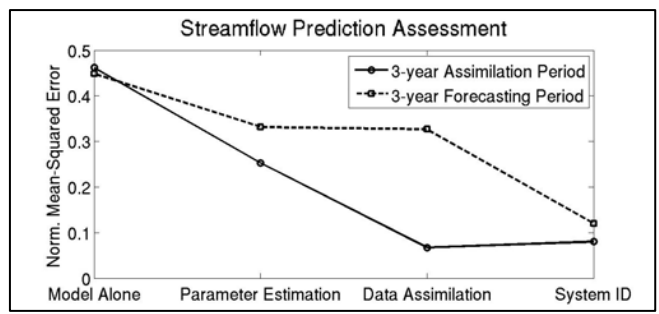

*Figure C.3: Squared error metrics from calibration and evaluation periods. Results show how system identification stores information from data assimilation into a predictive model by turning that model into a Bayesian network (like in **Figure C.2**) and then updating the network itself.*

The purpose of making these plots is first to locate (and compare the relative magnitudes of) individual deficiencies in the model. The story in this example is that the model is strongly underestimating the role of precipitation on the soil moisture storage variable, which suggests that the biggest weakness in this model is the infiltration function. Additionally we see model underestimations of influence between the soil moisture state and the surface and subsurface storage states, implying that soil moisture should play a larger role in mediating surface runoff and subsurface flow.

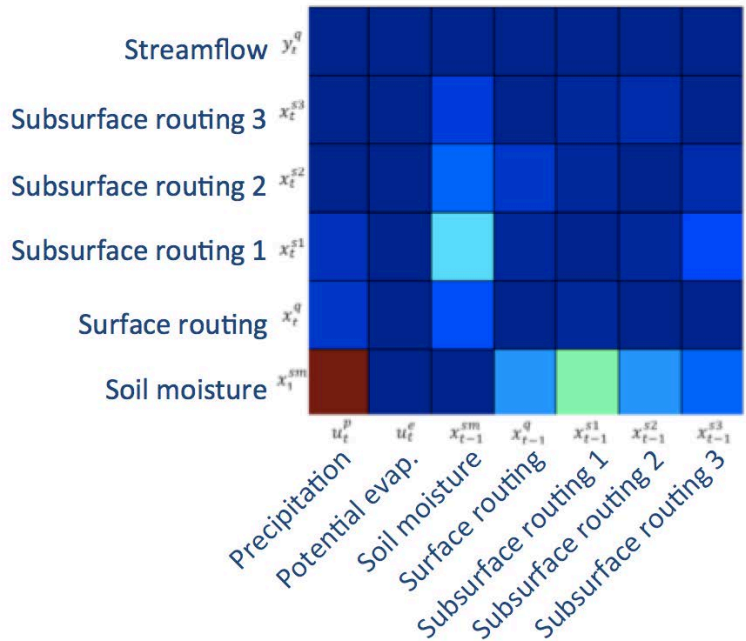

*Figure C.4*: *Absolute differences between pairwise directed information transfers in HyMod before vs. after assimilating daily streamflow observations over a period of three years. The degree of red shading reflects larger differences in process-level relationships indicated by observations.*